# CloudBrain-NMR: An Intelligent Cloud Computing Platform for NMR Spectroscopy Processing, Reconstruction and Analysis

Di Guo, Sijin Li, Jun Liu, Zhangren Tu, Tianyu Qiu, Jingjing Xu, Liubin Feng,
Donghai Lin, Qing Hong, Meijin Lin, Yanqin Lin and Xiaobo Qu*

*Abstract* — **Nuclear Magnetic Resonance (NMR) spectroscopy has served as a powerful analytical tool for studying molecular structure and dynamics in chemistry and biology. However, the processing of raw data acquired from NMR spectrometers and subsequent quantitative analysis involves various specialized tools, which necessitates comprehensive knowledge in programming and NMR. Particularly, the emerging deep learning tools is hard to be widely used in NMR due to the sophisticated setup of computation. Thus, NMR processing is not an easy task for chemist and biologists. In this work, we present CloudBrain-NMR, an intelligent online cloud computing platform designed for NMR data reading, processing, reconstruction, and quantitative analysis. The platform is conveniently accessed through a web browser, eliminating the need for any program installation on the user side. CloudBrain-NMR uses parallel computing with graphics processing units and central processing units, resulting in significantly shortened computation time. Furthermore, it incorporates state-of-the-art deep learning-based algorithms offering comprehensive functionalities that allow users to complete the entire processing procedure without relying on additional software. This platform has empowered NMR applications with advanced artificial intelligence processing. CloudBrain-NMR is openly accessible for free usage at https://csrc.xmu.edu.cn/CloudBrain.html.**

This work was partially supported by the National Natural Science Foundation of China (62371410, 62122064, 61971361 and 62331021), the Natural Science Foundation of Fujian Province of China (2021J011184 and 2023J02005), the President Fund of Xiamen University (20720220063), and Nanqiang Outstanding Talent Program of Xiamen University. (*Corresponding author: Xiaobo Qu, Email: quxiaobo@xmu.edu.cn)

D. Guo, S. Li and J. Liu are with School of Computer and Information Engineering, Fujian Engineering Research Center for Medical Data Mining and Application, Xiamen University of Technology, Xiamen 361024, China.

Z. Tu, T. Qiu, J. Xu, Y. Lin and X. Qu are with Department of Electronic Science, Fujian Provincial Key Laboratory of Plasma and Magnetic Resonance, Xiamen University, Xiamen 361005, China

L. Feng and D. Lin are with the Key Laboratory for Chemical Biology of Fujian Province, MOE Key Laboratory of Spectrochemical Analysis & Instrumentation, College of Chemistry and Chemical Engineering, Xiamen University, Xiamen 361005, China

M. Lin is with Department of Applied Marine Physics and Engineering, College of Ocean and Earth Sciences, Xiamen University, Xiamen 361102, China

Q. Hong is with the China Mobile Group, Xiamen 361005, China

## I. INTRODUCTION

**N**uclear Magnetic Resonance (NMR) spectroscopy, as an analytical technique that are widely used in biology [1]-[6], chemistry [7][8] and medicine [9], is a powerful tool that utilizes NMR phenomena to detect the composition and structure of molecules at the atomic level. The spectra are commonly presented in one-dimension or multi-dimension. The former may suffer from signal crowding, resulting in the overlapped spectral peaks and hard spectrum analysis. The latter alleviate the crowding but at the cost of significantly prolonged data acquisition time. To reduce this time, non-uniform sampling (NUS) can acquire partial data but need smart algorithms, e.g., Low-Rank (LR) [10] and deep learning (DL) [11][17][18], to fulfill the missing data. Thus, advanced algorithms may make NMR data processing more complex.

For example, the whole NMR data processing under NUS is summarized in Fig. 1. First, the raw NUS data collected from the spectrometer needs to be preprocessed using NMRPipe [12]. Second, reconstruction algorithms are employed to recover missing data and obtain high-quality spectra. Third, the data undergoes post-processing procedures. Finally, an analysis software is used for spectra visualization, peak-picking and quantitative analysis. The entire process requires switching between various software or programs, demanding deep knowledge on all programs and data conversion routines.

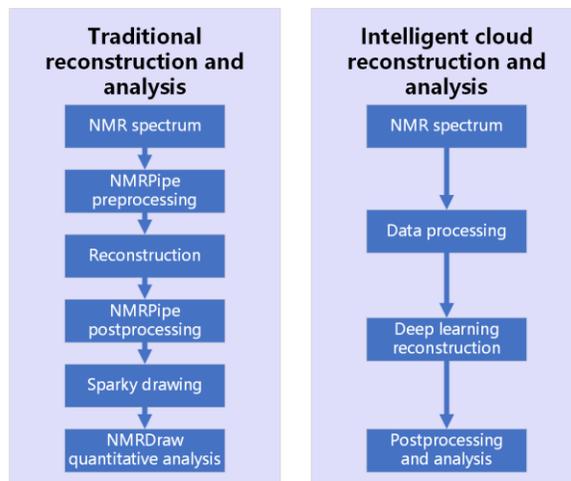

Fig. 1. Typical processing steps of the traditional way and the proposed platform.

Recently, researchers have designed integrated software or platforms for processing and analyzing NMR data, such as CCPN [13], NMRFx [14] and NMRBox [15]. The former two softwares are offline and need to be setup on personal computer or server. NMRBox supports remote desktop, allowing to install and use multiple professional data processing softwares such as NMRPipe. Yet, an online web-based processing and advanced artificial intelligence algorithms are missing. Thus, these platforms hinder the data flow integration and requires more user training.

NMR spectrometer vendors also provide processing and analysis softwares. For example, TopSpin [16] is a Bruker software used in a wide range of workflows. It has comprehensive functionalities and only needs few commands to process, display and analyze the NMR spectrum. But TopSpin requires software downloads and lacks the state-of-the-art artificial intelligence methods such as deep learning spectrum reconstruction [11][17][18].

Here, we will develop a one-site intelligent cloud computing NMR platform called CloudBrain-NMR. This platform is easy to operate and user-friendly, and integrates a series of spectrum reconstruction and analysis methods. It has three advantages:

1) *Multifunction:* CloudBrain-NMR provides rich functions, including data pre-processing, spectrum reconstruction, post-processing, generation of simulation data, neural network training, and spectrum analysis. It includes deep learning spectrum reconstruction, intelligent peak searching, peak height estimation, and quantitative analysis of molecular concentration.

2) *One-site processing:* The existing NMR analysis process requires complex tools and switching back and forth between various platforms and programs. Here, users only need to log in to the platform to complete all the process of spectrum, without the skills of programming or switching between multiple platforms or software.

3) *Fast and high-fidelity reconstruction:* The platform has implemented the state-of-the-art deep learning NMR approaches under one graphic processing unit and eight-core central processing units. These hardware enables fast spectrum reconstruction within one second [17].

## II. NMR Cloud Computing Platform

The complete workflow of NMR cloud platform is shown in Fig. 2 and the main steps include:

1) *Register and login:* The URL is https://csrc.xmu.edu.cn/CloudBrain.html [26] and the test **Account**: NMRTest1, **Password**: nmrtest1.

2) *Upload raw data and pre-processing:* Select the data type, sampling method, and pre-processing operations (adding window function, automatic phase correction, and intercept data range).

3) *Set NUS parameters:* Fill in NUS parameters if the data is fully sampled, otherwise skip this step if the data is undersampled.

4) *Spectrum Reconstruction:* Choose a reconstruction algorithm (LRHM, VIP and Modern) to reconstruct the spectrum.

5) *Postprocessing:* Perform the Fourier transform on the indirect dimension, correct the phase and remove the imaginary part. Also, users can view and download the reconstructed spectra.

6) *Peak picking:* Call the deep picker [22]-[24]. Set the minimum of peak intensity (default value is 5.5) and threshold of noise (default values is 3.0) to avoid the detection of noise. Also, users can view the results and download data.

7) *Generate dataset*: Generate synthetic data (training set) through the sum of simulated exponential functions. This data will be used to train the deep learning network since huge amount of real NMR is hard to acquire. This step can be skipped if the neural network has been trained.

8) *Train neural network:* Set the sampling rate of NUS, the number of training rounds, and select the training set. Notably, the sampling rate in the training set must match the target reconstructed data.

9) *Quantitative analysis:* Four steps are required for quantitative analysis. (a) Filling in the delta value to control the size of spectral peak window for peak area integration; (b) Determination of integration area by filling in value to observed region; (c) Click button to perform automatic integration; (d) Compute quantitative results.

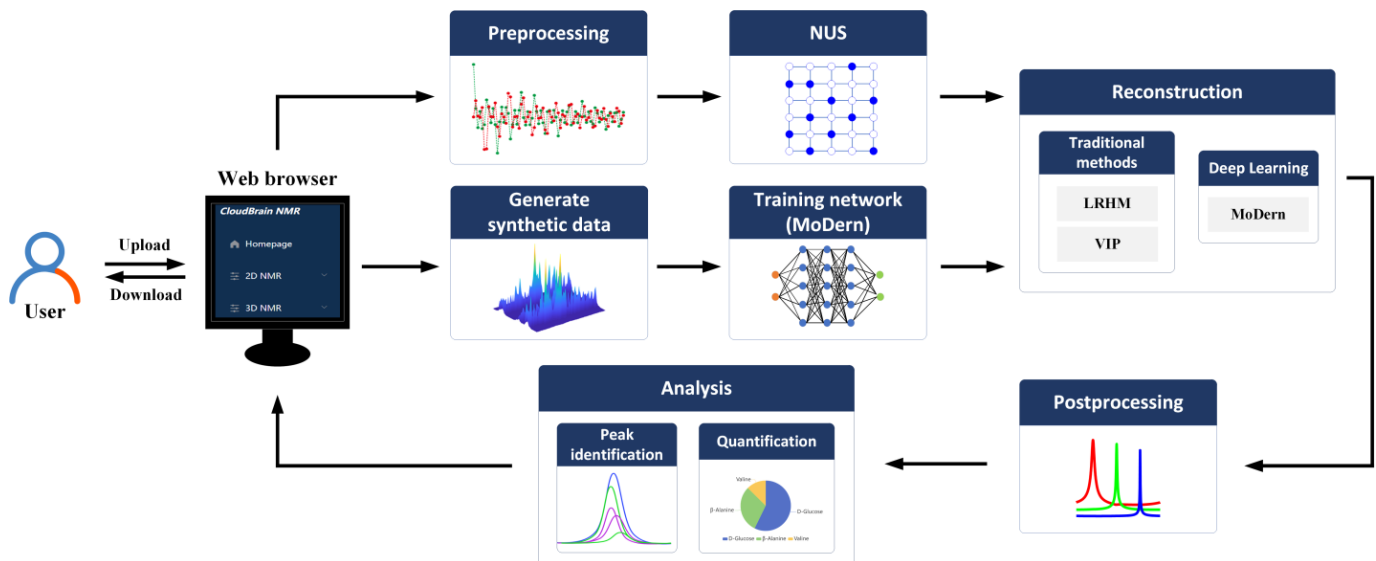

Fig. 2. Workflow of CloudBrain-NMR.

## III. SYSTEM FRAMEWORK

Fig. 3 shows the system architecture of the platform.

The four-tier system architecture based on browser/server mode is adopted: The frontend user interface layer, message queue (MQ) layer, backend server layer and data access layer (DAL). This architecture is beneficial to system development, maintenance and updating.

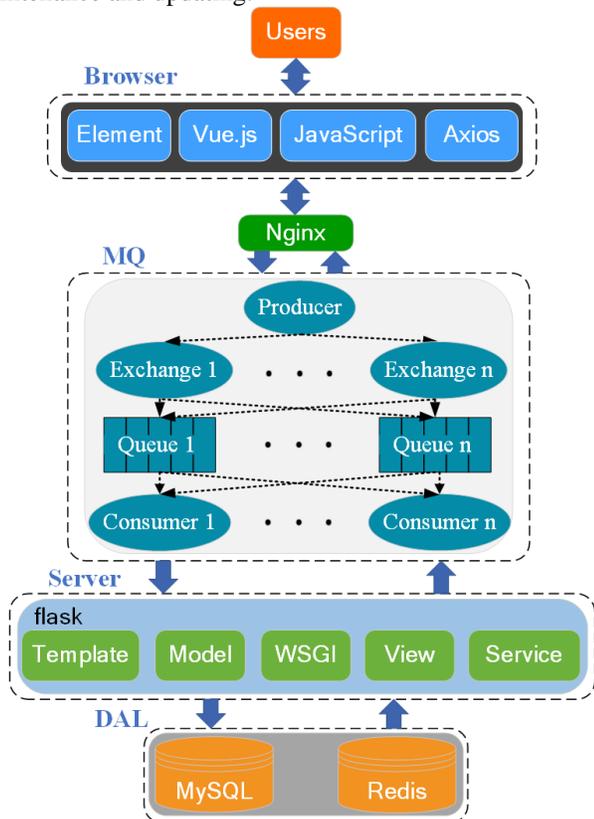

Fig. 3. System architecture of CloudBrain-NMR.

### A. Browser layer

The browser layer consists of user and frontend user interface parts. Components of the user interface include the Vue framework [28], Element component library [29], JavaScript [30], and Axios [31]. Vue is a JavaScript framework for building the CloudBrain-NMR interface, enabling dynamic interaction of webpage. The Element component adds functionality onto the webpage and Axios sends and processes requests, enabling interactions between the front and back ends and the database. With these components, one can easily access the platform through a web browser, triggering the encapsulated API to send the relevant request to the backend, and returning the visualized result in the browser from the server.

### B. Nginx

Nginx (Engine x) [27] is a high-performance HTTP and reverse proxy web server, achieving low memory usage, extremely fast startup, and strong concurrency capabilities. It provides load balancing, relieving the system pressure at high concurrency when multiple users access it simultaneously. Nginx can reduce user waiting time, improve efficiency, and ensure the stable operation of the platform. When processing multi-user usage on CloudBrain-NMR, the platform needs to handle requests of different functional modules. Then, Nginx can isolate module functions to prevent situations where some from taking too long to process and making others unavailable.

### C. Message queue layer

Message queue is a container for storing messages and placing transmitted data in the queue. This layer aims to ensure the high concurrency performance of the platform and control the system traffic, preventing system paralysis due to excessive traffic such that message loses. Meanwhile, asynchronous processing of messages is realized to accelerate the response and stability of the system. The platform adopts RabbitMQ [20] to prevent system crashes from data transmission between multiple layers and competition for server resources in high concurrency situation. It ensures the stability and availability of the platform when multiple users send requests simultaneously, and guarantees the concurrency pressure of more than ten people on a single server of the platform. For example, when multiple users call the reconstruction module simultaneously, the message queue adds the requests to the queue and processes the requests one by one in sequence to ensure normal function running.

### D. Server layer

The server layer receives all frontend requests, calls corresponding algorithms and computing resources, and provides fast computing services for user requests. This layer is mainly composed of a lightweight framework flask [32] that is written in python. The framework consists of built-in Web Server Gateway Interface (WSGI), view, model, template and service. When users submit data parameters on the web page, CloudBrain-NMR establishes a connection through WSGI, calls the view function to receive requests and parameters, and activate the corresponding functional modules to process the relevant data and interact with the database. Finally, the result is returned to the front end using the view function, and the relevant processing content is displayed to using the template. Service deploys all the algorithms used by the platform, which will be described in Section IV.

### E. Data access layer

The proposed platform adopts a data storage that integrates MySQL with Redis. MySQL is utilized to store structured data such as user registration information and spectrum processing results. Redis is applied to store email verification codes for CloudBrain-NMR user registration and password retrieval.

### F. Scheduling GPU to enable artificial intelligence

The platform runs on the heterogeneous computing graphics processing units (GPU) cloud server provided by China Mobile, and its system operating environment configuration is summarized in Table I. The entire project is managed and run through Docker container [33]. TensorFlow [34] is adopted to schedule GPU parallel computing to fast deep learning NMR training, target spectrum reconstruction and picking spectrum peaks. Requests on using GPU are lined in a queue manner, without worrying about resource competition, enabling the maximal utilization of the GPU.

## IV. DEPLOYED MACHINE LEARNING ALGORITHMS

This section introduces the core machine learning algorithms integrated on the cloud platform.

### A. Spectrum reconstruction with low-rankness or deep learning

To save the time of data acquisition in biological or chemical NMR experiments, only partial data are acquired from the spectrometer under NUS at the cost of introducing spectrum artifacts. To remove these artifacts and obtain a clear spectrum, a smart reconstruction algorithm is required to fulfill the missing data.

CloudBrain-NMR has three reconstruction algorithms, including two traditional methods (Low Rank Hankel Matrix (LRHM) [19] and VIrtual Peak (VIP) [18]) and one state-of-the-art deep leaning method (MoDern [17]).

LRHM method [10] exploits the low-rank Hankel property of Free Induction Decay (FID). Through modeling the FID as the sum of several exponentials and converting the FID into Hankel matrix, the rank of this matrix equals to the number of exponentials [19]. This rank will be small if the number of exponentials is much smaller than the data length of the FID and this prior can be used to regularize the reconstructed spectrum [19].

However, LRHM tends to distort small spectral peaks if the acquired FID data points are very limited. To address this issue, the VIP method incorporates extra prior information of spectral peaks [18], such as center frequency and shape of the spectral peak, into the reconstruction model through self-learning subspaces. This strategy achieves the high-fidelity reconstruction, particularly on the reconstruction of low intensity peaks, and significantly improves the accuracy of quantification, including the distances between nuclear pairs, and the concentration of metabolites in mixture [18].

TABLE I
SYSTEM OPERATING ENVIRONMENT CONFIGURATION

| Name | Function |
| --- | --- |
| CentOS 8 | Operating system |
| Intel Xeon Gold 6240@2.6GHz | Processor model/Main frequency |
| 8-Core CPU, 64G RAM | System operation hardware |
| NVIDIA T4 | Neural network training graphics card |
| CUDA 11.7 | Universal parallel computing architecture |
| Driver 515.76 | Video driver |
| 5Gbit/s | Network bandwidth |
| Python 3.6.13 | Code compiler |
| Docker | Container |
| RabbitMQ | Message queue middleware |
| MySQL5.7, Redis5.0 | Database |
| Nginx | HTTP and reverse proxy server |
| TensorFlow | Deep learning framework |

Recently, deep learning has received extensive attention and has been applied to the fields such as biomedicine and chemistry. MoDern is a sparse inspired meta learning network that can handle mismatch between training and target NMR data, enabling ultra-fast high-fidelity reconstruction [17]. Under the principle of meta-learning, MoDern defines an optimal threshold to generalize the network to robustly reconstruction spectrum under multiple acceleration factors of fast data acquisition. Spectrum artifacts are gradually removed in network, and finally, a high-quality spectrum is output.

### B. Peak identification with DEEP Picker

To accurately identify the spectral peak position and extract other spectral information, DEEP Picker [22]-[24] proposed by Li *et al* deconvolves and picks spectral peaks with deep learning [22]-[24]. The network is trained with many simulated synthetic spectra consisting of known composition with different degrees of crowdedness. An advanced function is its powerful capability of correctly identifying overlapping peaks, which are always challenging to existing computational methods and even professional spectroscopists.

### V. IMPLEMENTATION AND RESULTS

This section describes all functional modules on cloud.

### A. Raw data uploading and pre-processing module

In this module, the raw FID data is uploaded and read by the functions that are extracted from an open-source code nmrglue [21], which is compatible to two representative vendors (Bruker and Varian). The sampling schemes (full sampling and NUS) should be set first and the uploaded data will be processed accordingly. Then, performing the common pre-data processing steps, including sine windowing (as shown in Equation (1) [12]), zero padding, Fourier transform, phase correction, imaginary part removal, and intercept the spectral region of interest. The sine window function is defined as

$$f(x_i) = \sin(\pi \times a + \pi \times \frac{(b-a) \times i}{s-1})^p , \qquad (1)$$

where $a$ and $b$ specify the starting and the ending points of the sine-bell in units of $\pi$ radians, respectively; $p$ indicates the exponent of the sine-bell; $s$ represents the number of points in the window function. Default values of $a$, $b$ and $p$ are 0.0, 1.0, and 1.0, respectively. Notably, p could be non-integer.

The extraction function preserves the spectral region of interests and removes the rest part of spectrum. Extraction is performed on the direct dimension (fully sampled dimension) of the 2D or 3D spectrum. This processing is not mandatory but it helps excluding the interference signals from non-interest spectrum interval and saving the computation and storage resources. The interface of pre-processing module is shown in Fig. 4.

### B. NUS module

This module is not necessary since the NUS scheme could be automatically read from the raw FID data if the NUS has been conducted physically on the spectrometer. If the FID data is fully sampled, then this module is useful to simulate the NUS scheme. On our platform, the Poisson NUS is simulated to obtain partial FID data, which will be used for the subsequent reconstruction. The Poisson distribution is chosen because it efficiently captures exponential signal characteristics, particularly in regions with rapid signal variations, and may improve the sensitivity [35].

### C. Reconstruction module

Reconstruction is to recover the missing FID data points and then obtain a clear spectrum when NUS is applied. For 2D NMR, MoDern [17], VIP [18] and LRHM [19] have been deployed on the platform. For 3D NMR, only MoDern [17] is deployed.

Regarding the computation time, the deep learning method (MoDern) runs much faster due to the powerful GPU and the non-iterative nature of a trained neural network. The LRHM and VIP use eight-core CPU parallel computing to save significant amount of reconstruction time. Taking the compuation time for 2D NMR as an example (Table II), MoDern runs ultrafast (within 1 second) and LRHM requires 17~40 seconds.

TABLE II
COMPARISON OF RECONSTRUCTION TIME ON CLOUD

| Sample | Spectra | | Reconstruction time on cloud (s) | | |
| --- | --- | --- | --- | --- | --- |
| | Type | Size | MoDern | VIP | LRHM |
| Ubiquitin | HSQC | 1024×98 | 0.39 | 240.61 | 38.35 |
| CD79b | HSQC | 116×256 | 0.38 | 120.57 | 17.75 |
| Gb1 | HSQC | 1466×170 | 0.43 | 293.49 | 40.25 |
| Ubiquitin | TROSY | 512×128 | 0.41 | 167.58 | 25.20 |

Note: The spectrum size is $p \times q$ where $p$ is size of the direct dimension, whose acquisition is very fast and usually fully sampled, and $q$ is the size of indirect dimension, whose acquisition is slow and is undersampled (25% data are sampled here).

To verify the quality of a reconstructed spectrum, the Pearson correlation coefficient is adopted to measure the correlation between a reconstructed spectrum $\hat{x}$ and its fully sampled one $x$ according to Equation (2):

$$R^2(\hat{x}, x) = \left( \frac{cov(\hat{x}, x)}{\sigma_{\hat{x}} \sigma_x} \right)^2 , \qquad (2)$$

where $cov(\bullet)$ and $\sigma$ denotes the covariance and standard deviation, respectively.

Taking a 2D NMR reconstruction as an example (Fig. 5), both LRHM and MoDern provides high-fidelity reconstructions, achieving a higher Pearson correlation coefficient $R^2$ than 0.999. For the 3D NMR (Fig. 6), MoDern can finish the reconstruction in 12.12 seconds.

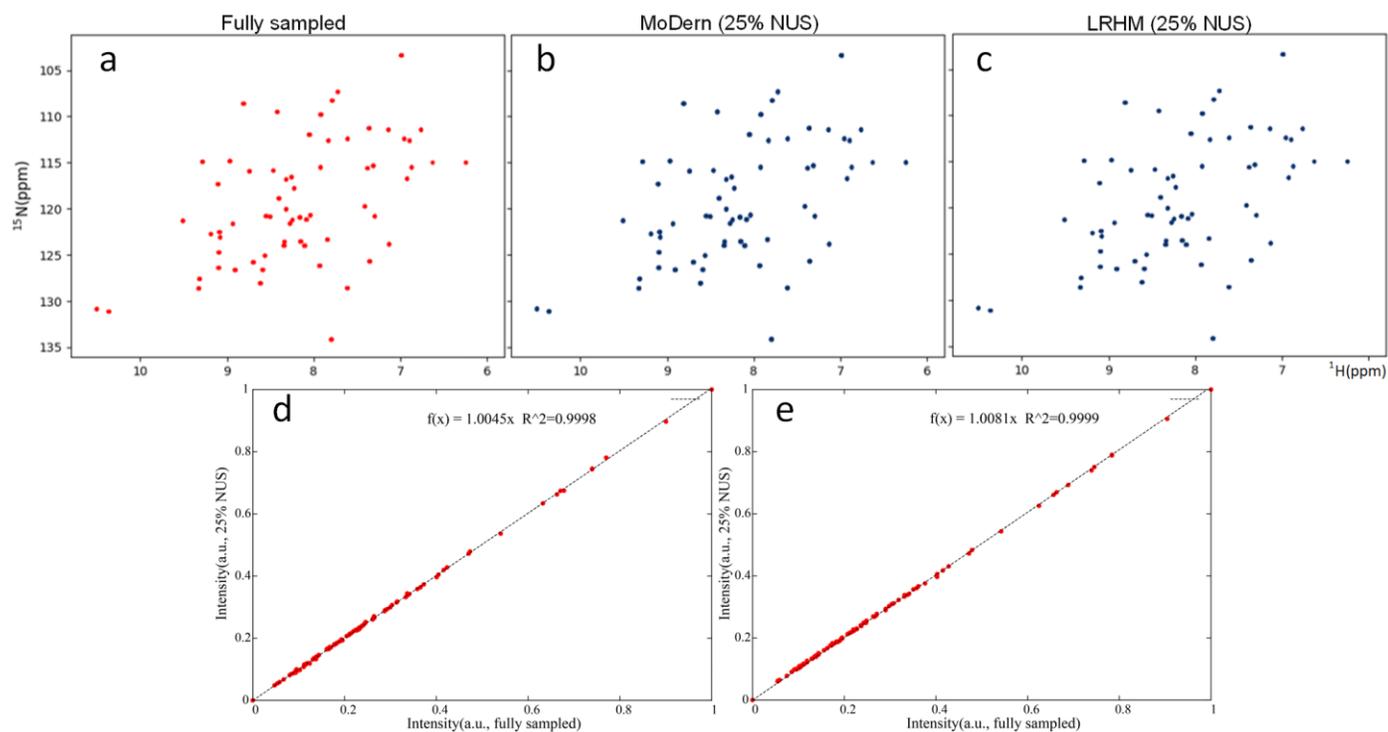

Fig. 4. Pre-processing interface on the CloudBrain-NMR.

Fig. 5. A reconstruction example of the $^1$H-$^{15}$N HSQC 2D spectrum of a protein Gb1 on our platform. (a) is the fully sampled spectrum, (b) and (c) are the reconstructed spectrum from 25% NUS data by the deep learning method (MoDern) and the low rank Hankel matrix method (LRHM), respectively, (d) and (e) are the correlations of spectral peaks between the fully sampled spectrum and the reconstructed one using MoDern and LRHM, respectively.

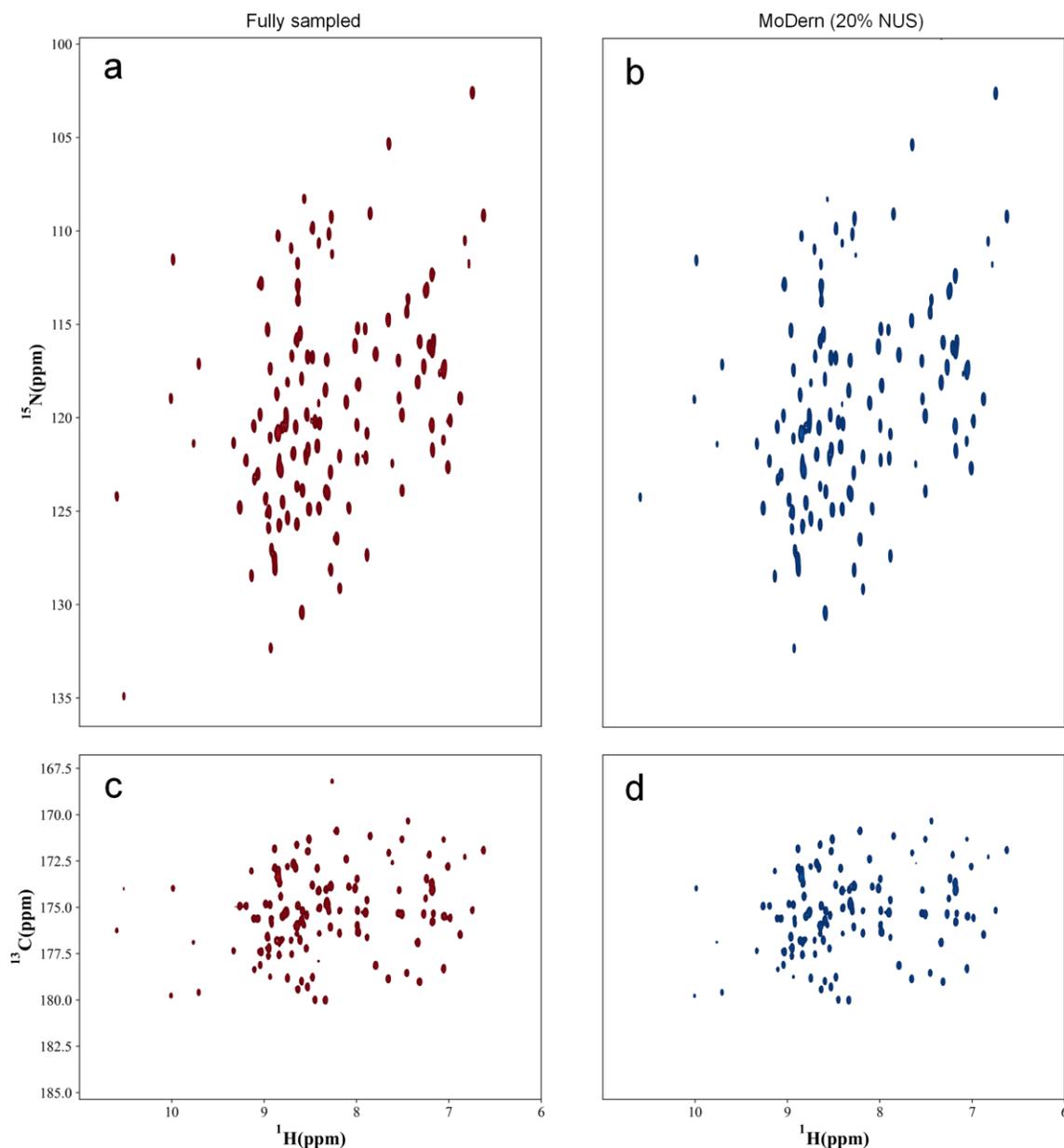

Fig. 6. A reconstruction example of the HNCO 3D spectrum of an azurin protein on our platform. (a) and (c) are projections on $^1$H-$^{15}$N and $^1$H-$^{13}$C planes of the fully sampled referenced spectrum. (b) and (d) are projections on $^1$H-$^{15}$N and $^1$H-$^{13}$C planes of the reconstructed spectrum.

### D. Post-processing module

This interface of post-processing module is shown in Fig. 7 and the workflow on cloud is summarized in Fig. 8. This module mainly processes the indirect dimension of spectral data. The indirect dimension is 1D vector (or 2D plane) for 2D (or 3D) NMR.

Taking 3D as an example, two indirect dimensions need to be processed, and the post-processing operations include: Sine windowing according to Equation (1), Fourier transform, automatic phase correction and imaginary part removal. Among them, the Fourier transform has a total of five modes ('Default', 'Auto', 'Alternative', 'Inverse', 'Negate'), how to choose these modes depends on the specific experimental conditions and data quality, more details can be found in nmrglue [21].

Finally, spectrum data in *.ft2 format is saved or can be downloaded.

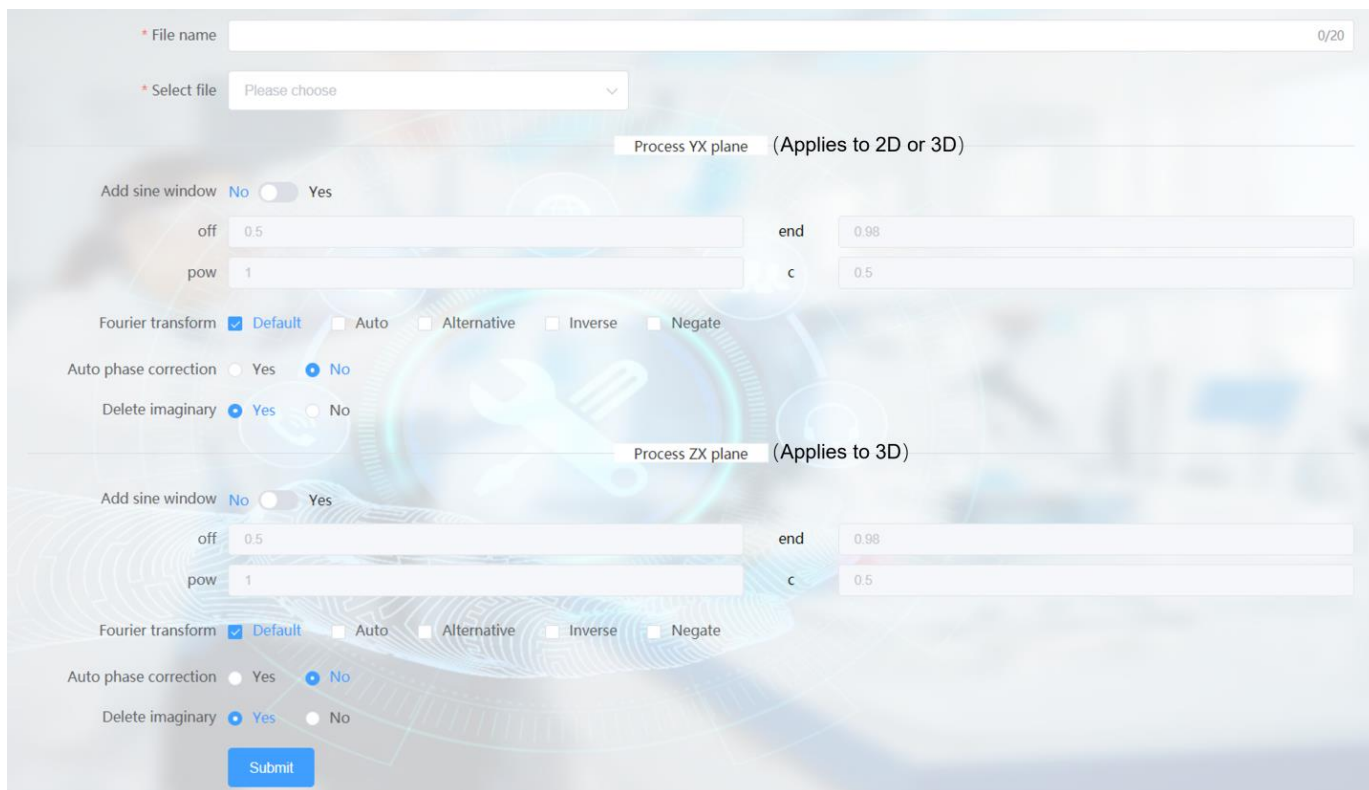

Fig. 7. Post-processing interface on the CloudBrain-NMR.

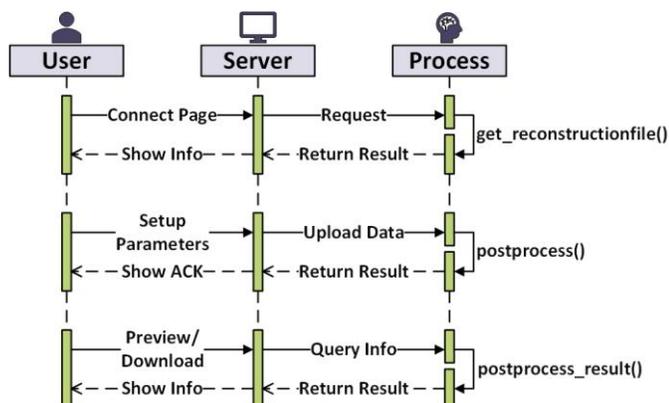

Fig. 8. The workflow of post-processing on the cloud.

*E. Peak identification module*

The module is designed to identify spectral peaks and output peak information. A state-of-the-art deep learning approach, DEEP Picker [22]-[24], is adopted for identification. The workflow and interface of peak identification is provided in Fig. 9 and Fig. 10, respectively.

On this page, one needs to firstly set the minimal peak intensity scale, noise threshold, and chemical sample type (including protein and metabolite). Then, spectrum data (*.ft2 format) is selected from the online database or local files. By clicking the "Submit", parameters and data will be sent from the user to server. Next, deep picker on cloud will be called to identify peaks and mine peak information. Finally, processed results will be saved on server and sent back once the query is received.

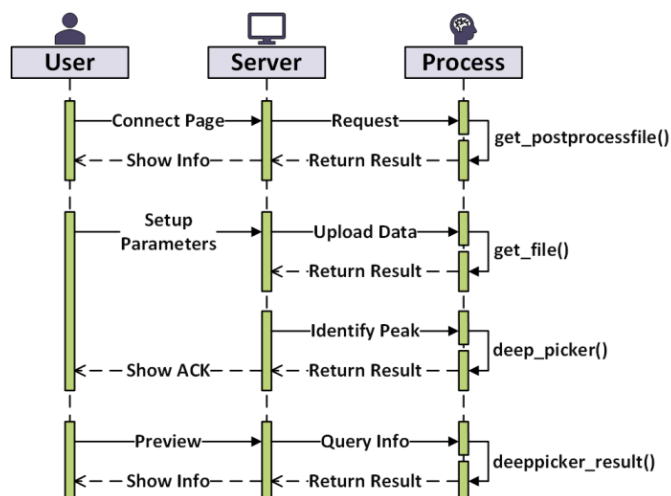

Fig. 9. The workflow of peak identification on the cloud.

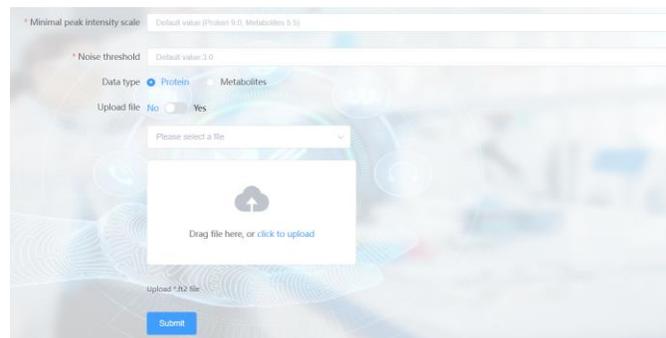

Fig. 10. Peak identification interface on the CloudBrain-NMR.

An identification example is tested on a 2D NMR spectrum (Fig. 11). Most peaks have been identified and marked with red cross symbols (Fig. 11(a)). To further measure the correctness of identification, a confidence is defined as the score according to

$$\text{Softmax}\left(z_p\right) = \frac{exp\left(z_p\right)}{\sum_{q=1}^{Q} exp\left(z_q\right)}, \qquad (3)$$

where z represents the data points with the same chemical shift, $p$ represents the $p^{\text{th}}$ point in z, $Q$ is the total number of data points, and $q$ represents the $q^{\text{th}}$ ($q$=1,2,...,$Q$) spectral peak. Fig. 11(b) shows that more than half peaks have a confidence that is greater than 0.9 while the rest ones has a confidence between 0.6~0.9.

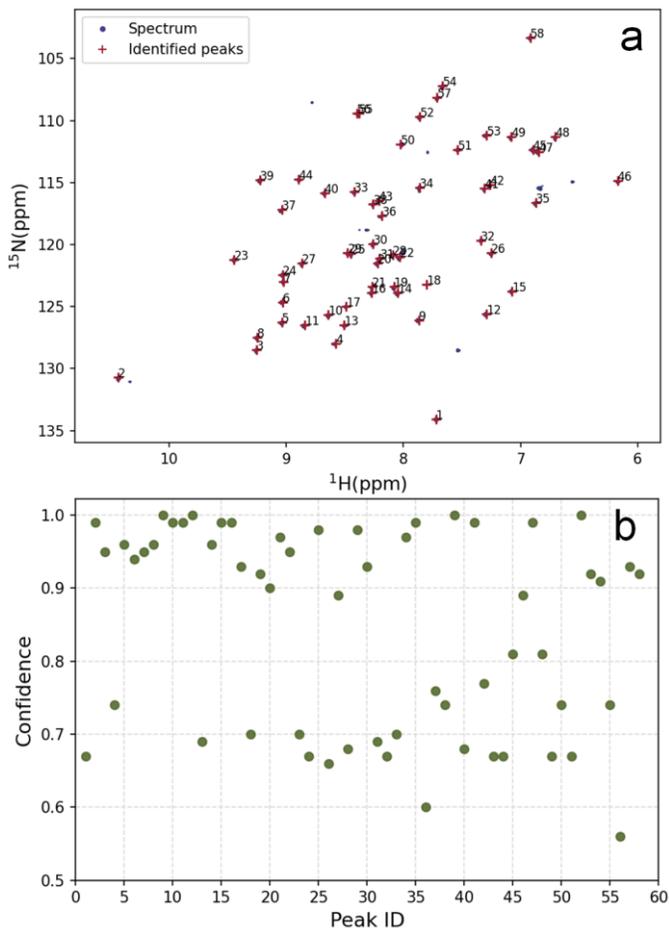

Fig. 11. Peak identification test on a 2D $^1$H-$^{15}$N spectrum of a protein Gb1. (a) Identified peaks are marked with red cross symbol, (b) confidence distribution of the identification. Note: The range of 5.8-10.8 ppm of the full spectrum is intercepted for test.

### F. Generating simulation data module

The module simulates FID signal to build a training set for deep learning spectrum reconstruction. The FID signal is modeled as a superposition of a finite number of exponential functions [25] according to

$$y_m = \sum_{j=1}^{J} \left(A_j e^{i\phi_j}\right) e^{-\frac{m\Delta t}{\tau_j}} e^{im\Delta t\omega_j}, \qquad (4)$$

where $i$ denotes the imaginary unit, $\Delta t$ denotes the time interval between two sampling points [25], $y_m$ represents the $m^{\text{th}}$ ($m$=1,2,..,$M$) sampled FID data point, $A_j$, $\phi_j$, $\tau_j$, $\omega_j$ indicate the amplitude, phase, damping factor, and angular frequency of the $j^{\text{th}}$ ($j$=1,2,..,$J$) exponentials, respectively. The sampling rate of NUS is defined as the ratio of $M/N$ where $N$ is the number of fully sampled FID data points. In simulation, spectral parameters in Equation (4) is chosen randomly from a uniform distribution [11] and a sampling rate has to be set. For example, to train the MoDern, 4000 data samples are generated under a sampling rate of 25%.

### G. Neural network training module

Network training is deployed on cloud, meaning that users do not need to buy graphics processing unit. Through visit this module on website, users can customize parameters, e.g., the sampling rate, to match the application to obtain the best reconstructed spectrum. Even with the mismatched sampling rate between the training and target spectra, this limitation can be overcome well with the state-of-the-art deep learning spectrum reconstruction algorithm, MoDern, which achieves high-fidelity reconstruction [17]. Training usually needs 3.9 hours for 2D NMR and 15.1 hours for 3D NMR. The training process can be skipped since exponential functions have been synthetized to train a general solution for NMR spectrum reconstruction [11][17].

### H. Quantitative module

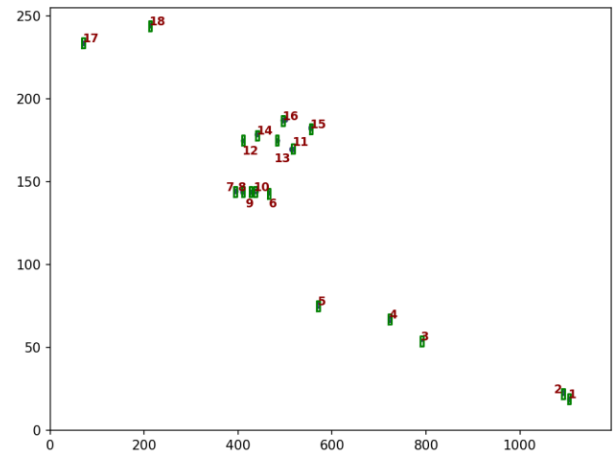

Fig. 12. Peak assignment of a $^1$H-$^{13}$C HSQC spectrum.

This module first marks each spectral peak in a range of intercepted region with automatic numbers (e.g. peak 1-18 in Fig. 12). Then, a parameter "Delta" shoud be set to determine the integration range of spectral peaks (the size of black dashed box in Fig. 13).

For a specific peak, zooming is to fill in the spectral peak number and range for viewing. Fig. 13 shows the spectral peak regions under different parameter settings. According to the initial set values, observe the area of spectral peak. If the important information of the spectral peak is not included in the area, or the spectral peaks are not fully displayed, the user can adjust the paddings of the X-axis and Y-axis to view the full spectral peaks. And Delta value can adjust the size of the window of spectral peak for peak area integration. The larger the values of padding, the smaller the amplification factor of the spectral peak. Peak identities and their integration values will be automatically saved online.

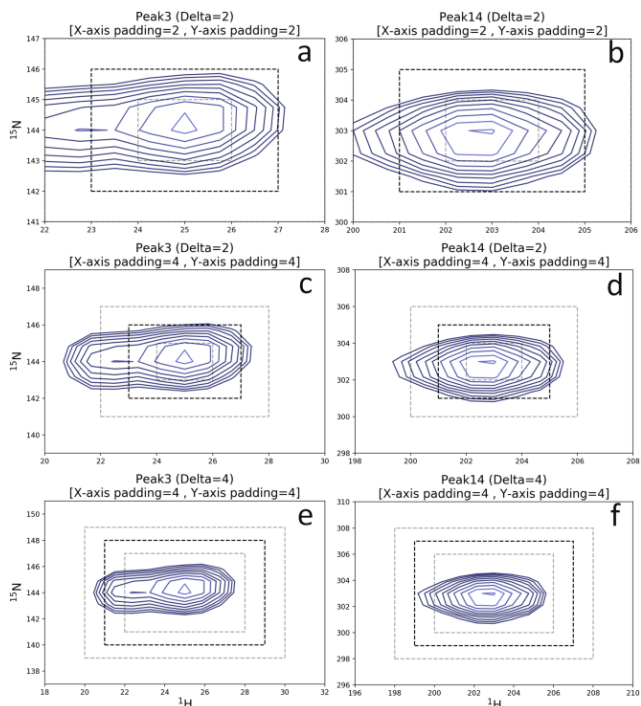

Fig. 13. Integration regions of different slices of 2D $^1$H-$^{13}$C HSQC spectrum of the mixture. Note: The black dashed line indicates the integration region. (a), (c) and (e) shows different padding values and Delta values for peak #3 in Fig. 12. (b), (d) and (f) shows different padding values and Delta values for peak #14 in Fig. 12.

## VI. Conclusions

In this work, we have developed CloudBrain-NMR, an intelligent cloud computing platform to process, reconstruct and analyze NMR spectroscopy. Notable deep learning functions, such as undersampled spectrum reconstruction and automatic peak picker, have been integrated. CloudBrain-NMR is an open-access platform at https://csrc.xmu.edu.cn/CloudBrain.html. Users only need to visit website through browser and do not need to install any software. Simultaneous visit for multiple users is supported, which has been found very useful in biochemical training courses, such as BioNMR Advanced Tools seminar held at University of Gothenburg [36][37]. In the future, we plan to enhance the CloudBrain-NMR by integrating other state-of-the-art artificial intelligence functions and provide reliable services for the NMR community.

## Acknowledgments

The authors thank Drs. Dawei Li and Rafael Brüschweiler (The Ohio State University) for providing the DEEP Picker code, Jonathan J. Helmus and Christopher P. Jaroniec for the nmrglue data processing code and Prof. Vladislav Orekhov (University of Gothenburg) for valuable suggestions.